# Large Magnetocaloric Effect and Adiabatic Temperature Change in the Tellurate: $Gd_2TeO_6$

Abhijit Nayak[1], Gurunathan Angusamy[2], Kusampal Yadav[3], S. K. Panda[4], Devajyoti Mukherjee[3], Raivo Stern[2], Tanmoy Chakrabarty[5], and Krishnamurthy Jyothinagaram[1*]

[1]*Department of Physics, National Institute of Technology, Andhra Pradesh, Tadepalligudem, Andhra Pradesh 534101, India*

[2]*National Institute of Chemical Physics and Biophysics, Tallinn 12618, Estonia*

[3]*School of Physical Sciences, Indian Association for the Cultivation of Science, Kolkata, West Bengal 700032, India*

[4]*Department of Physics, Bennet University, Greater Noida, Uttar Pradesh 201310, India*

[5]*Division of Sciences, Krea University, Sri City, Andhra Pradesh 517646, India*

## Abstract

We investigated the structural, magnetic, thermodynamic, and magnetocaloric properties of polycrystalline $Gd_2TeO_6$, a rare-earth tellurate that has remained unexplored from the perspective of low-temperature magnetism and magnetic refrigeration. Powder X-ray diffraction confirms the formation of single-phase orthorhombic $Gd_2TeO_6$ with space group $P2_12_12_1$, containing two crystallographically distinct seven-coordinate Gd sites. Temperature-dependent magnetization measurements reveal no conventional long-range magnetic ordering down to 2 K, while a low-field irreversibility develops below approximately 4 K. Curie-Weiss analysis yields $\theta_{CW}$ of -1.25 K, indicating weak predominantly antiferromagnetic interactions and confirming the dominant contribution of the two $Gd^{3+}$ ions. The magnetization reaches approximately 97 % of the theoretical saturation moment at 7 T and exhibits negligible hysteresis. Field-dependent heat-capacity measurements show no sharp zero-field anomaly above 2 K, while the field-induced broad maximum is consistent with a redistribution of low-energy magnetic entropy. A Debye-Einstein analysis was used to separate the lattice and magnetic contributions to the heat capacity. The resulting magnetocaloric response is substantial, with $-\Delta S_M$ = 21.40, 48.51, and 54.81 J kg$^{-1}$ K$^{-1}$ for field changes of 0-2, 0-5, and 0-7 T, respectively, at the lowest measured temperature of 2.5 K. The corresponding maximum volumetric entropy change reaches 400.50 m J cm$^{-3}$ K$^{-1}$ for 0-7 T. Extending the magnetization measurements to 14 T yields a maximum measured $-\Delta S_M$ of 63.51 J kg$^{-1}$ K$^{-1}$, approaching the theoretical entropy limit of 64.25 J kg$^{-1}$K$^{-1}$ for two $Gd^{3+}$ ions per formula unit. An estimation of adiabatic temperature change ($\Delta T_{ad}$) gives approximately 22.7 K of temperature change for 0-7 T. First-principles calculations identify closely competing antiferromagnetic states, with GGA + $U$ favoring antiparallel alignment between two sublattices of internally parallel Gd moments. Combined experimental and theoretical results highlight $Gd_2TeO_6$ as a promising Gd-based oxide for cryogenic magnetocaloric applications.

***Corresponding Author's Email ID*:** krishnamurthy@nitandhra.ac.in

## 1. Introduction:

The increasing demand for reliable refrigeration at cryogenic and sub-kelvin temperatures has driven considerable interest in alternative cooling technologies capable of operating at extremely low temperatures. Such temperatures are essential for a variety of applications, including quantum technologies, ultrasensitive detectors, and low-temperature condensed-matter experiments [1–4]. Dilution refrigerators based on the $^{3}He$-$^{4}He$ mixture are among the established technologies for reaching the millikelvin temperature range [5]. However, the limited availability accompanied by the cost of $^{3}He$, whose major supply source is associated with the radioactive decay of tritium, has motivated the exploration of alternative refrigeration technologies [6–8]. Adiabatic demagnetization refrigeration (ADR), based on the magnetocaloric effect (MCE), is one such route in which the temperature of a magnetic refrigerant is reduced through adiabatic demagnetization [9–11] .

The performance of an ADR refrigerant is determined by the availability of magnetic entropy at low temperatures, the strength of magnetic interactions, the magnetic ordering temperature, and the reversibility of the magnetization process [12,13]. In this context, $Gd^{3+}$ has beomme as one of the most extensively investigated magnetic ions for cryogenic refrigeration owing to its half-filled $4f^7$ configuration, corresponding to $S = 7/2$, $L = 0$, and $J = 7/2$ [14,15]. The large spin provides a high magnetic entropy, while the absence of orbital angular momentum reduces single-ion crystal-field effects to a large extent. As a result, Gd-based compounds have been widely explored as low-temperature magnetocaloric and ADR materials [16].

Recent investigations show the continuing interest in Gd-based oxides for cryogenic magnetic refrigeration. For example, $Gd_3BSi_2O_{10}$ was reported to exhibit a large cryogenic magnetocaloric response, which was associated with its high magnetic-ion density and relatively weak magnetic interactions [17]. Recently, the Gd-rich silicate $NaGd_9(SiO_4)_6O_2$ demonstrated a magnetic entropy change of approximately 60 J kg$^{-1}$ K$^{-1}$ under a 0-7 T field change at 2.6 K, together with a volumetric entropy change of 382.8 mJ cm$^{-3}$ K$^{-1}$, highlighting the importance of both gravimetric and volumetric performance for compact cryogenic refrigeration systems [18]. $Gd_4Ga_2O_9$ has also recently been investigated as a cryogenic magnetocaloric oxide, demonstrating a maximum entropy change of 25.77 J kg$^{-1}$ K$^{-1}$ under 0-7 T [19].

Importantly, recent research has increasingly focused not only on maximizing the magnetic entropy change but also on suppressing magnetic ordering to lower temperatures [20,21]. A recent review of frustrated Gd and Yb-based oxides further highlights that magnetic frustration and competing interactions can retain substantial entropy at temperatures well below the characteristic magnetic interaction scale, making such materials promising for mK ADR [22].

The importance of these principles has also been demonstrated through direct refrigeration experiments. $GdCrTiO_5$, an exchange-frustrated Gd-based oxide, was investigated using the direct ADR method and reached minimum temperatures of 0.4, 0.67 and 0.88 K starting from 1.8, 2.7 and 3.7 K, respectively [23]. More recently, $Gd_2B_2MoO_9$ was demonstrated as a frustrated Gd-based magnetic refrigerant capable of reaching 0.16 K using a quasi-adiabatic demagnetization apparatus. The material exhibited entropy changes of 13.3, 33.7 and 45.1 J $kg^{-1}$ $K^{-1}$ for field changes of 0-1, 0-2 and 0-3 T, respectively, emphasizing the scope of frustrated Gd-based oxides for ultralow-temperature refrigeration [16].

These recent findings indicate that an effective cryogenic refrigerant requires not only a large density of magnetic moments but also sufficiently weak magnetic interactions to suppress the magnetic entropy from being utilized by long-range ordering at the target operating temperature. In this regard, chemical design of Gd-based oxides provides a promising route for balancing magnetic entropy, interaction strength, structural stability, and volumetric cooling capacity. Recent work has additionally demonstrated strategies in which the concentration of Gd is combined with other magnetic ions to simultaneously manipulate the low-temperature ordering temperature and magnetic entropy. For example, Gd-Yb fluoride compounds have been investigated with the direct objective of obtaining large magnetic entropy changes while maintaining very low ordering temperatures [24].

Within this broader context, rare-earth tellurates of the $RE_2TeO_6$ family represent an interesting but comparatively unexplored class of materials from the perspective of magnetism as well as magnetocaloric refrigeration. The crystal structure of $Gd_2TeO_6$ was established in 2003, revealing an orthorhombic structure with space group $P2_12_12_1$ containing two crystallographically independent seven-coordinate $Gd^{3+}$ sites and isolated TeO6 octahedra [25]. Later investigations on $RE_2TeO_6$ compounds have mainly focused on their synthesis, vibrational properties, luminescence, and optical applications [26–28]. Thus, despite the presence of two magnetic $Gd^{3+}$ ions per formula unit, a systematic investigation of the magnetic and magnetocaloric properties of $Gd_2TeO_6$ remains lacking, to the best of our knowledge.

The large theoretical magnetic entropy associated with $Gd^{3+}$ further strengthens the motivation for investigating $Gd_2TeO_6$ as a cryogenic magnetocaloric material. For two $Gd^{3+}$ ions per formula unit, the maximum magnetic entropy is given by $S_{mag}$ = 2Rln(2$J$ + 1) = 2Rln8 = 34.57 J $mol^{-1}K^{-1}$ per unit mass, the corresponding theoretical entropy for $Gd_2TeO_6$ is approximately 64.25 J $kg^{-1}$ $K^{-1}$. For comparison, the isostructural $Gd_2WO_6$ has a lower theoretical value of approximately 58.2 $kg^{-1}$ $K^{-1}$, owing to its higher molar mass, and exhibits an experimentally determined maximum magnetic-entropy change of 34.8 $kg^{-1}$ $K^{-1}$ for a field change of 0-7 T at 4 K [29]. Thus, the lower molar mass of $Gd_2TeO_6$, together with its two $Gd^{3+}$ magnetic sublattices and the possibility of weak magnetic interactions at low temperatures, provides a strong motivation for investigating its magnetocaloric properties for cryogenic refrigeration.

In the present work, polycrystalline $Gd_2TeO_6$ was synthesized by the conventional solid-state reaction method, followed by detailed structural, magnetic, thermodynamic, and magnetocaloric characterization. The magnetic properties were investigated through temperature and field-dependent magnetization measurements, while the thermodynamic response was examined using field-dependent heat capacity measurements. The magnetocaloric performance was evaluated in terms of the magnetic entropy change, volumetric entropy change, and adiabatic temperature change using magnetization and heat capacity approaches. In addition, first-principles calculations were performed to investigate the electronic structure and competing magnetic configurations of $Gd_2TeO_6$.

## 2. Experimental Characterization and DFT Methodology Details

### 2.1. Sample preparation and characterization details

The polycrystalline sample of $Gd_2TeO_6$ was synthesized by the standard solid-state reaction method by reacting $Gd_2O_3$ and $TeO_2$. Initially, the stoichiometric amount of $Gd_2O_3$ and $TeO_2$ was hand-ground thoroughly, before which the rare-earth oxide was pre-calcinated at 900°C for 12 hours to eliminate the water and carbonates. The resulting mixture was heated at 550°C for 12 hours, the resulting powder was then pressed into discs of 15 mm diameter with a pressure of 7 tons and the resulting discs were sintered at 700°C and 900°C for 12 hours with one intermediate grinding. For all the heat treatments, the heating and cooling rate of 5°C was followed [26].

For the investigation of phase purity of the sample, powder X-ray diffraction (PXRD) was performed on approximately 1 g of the sample in the Malvern PANalytical XRD equipment in the 2ϴ range of 10° to 90° with the step size of 0.01313°. Temperature-dependent magnetic measurements were performed on the Magnetic Property Measurement System (MPMS3, Quantum Design USA), whereas, for the field-dependent magnetic properties both the MPMS and Physical Property Measurement System (PPMS) (Quantum Design, USA) were used. Similarly for the Specific heat measurement, PPMS was used where the heat capacity of the sample was assessed by the thermal relaxation method.

### 2.2.Theoretical Calculation Details

To calculate the electronic and magnetic ground state properties of $Gd_2TeO_6$, the Spin-polarized DFT method was implemented in the VASP 5.4.4 package, with the projected augmented wave (PAW) method. The plane wave cutoff of 650 eV was considered, and the K-mesh sampling was performed in 6*4*3 Monkhorst-Pack grid. For the structural relaxation the energy cutoff was fixed to be $10^{-6}$ eV, and the relaxation was performed with the conjugate gradient algorithm until the inter-atomic forces were reduced to 0.005 eV/Å. Whereas, for the Self-consistent field (SCF) and the non Self-consistent field (NSCF) calculations, the energy cut-off criteria was set to be $10^{-8}$ eV. Furthermore, to take into account the localized $4f$ electrons of $Gd^{3+}$, the Hubbard's $U$ value correction of $U$ = 5, 6, 7, and 8 eV was used in the case of the Gd atoms to verify the consistency of the calculations. To take into account the $f$-electrons as valence electrons for the Gd atoms, the appropriate pseudopotential for Gd atoms was considered, where the Valence-elctron configuration, $4f^{7.5}5S^2 5p^6 5d^{0.5} 6S^2$ was used, and for Te and O, the pseudopotentials possessing the valence electron configuration of $5s^2 5p^4$, and $2S^2 2p^4$ were used.

## 3. Results and Discussion

### 3.1. Structural Characterization:

**Fig. 1.** depicts the powder X-ray diffraction pattern of $Gd_2TeO_6$ measured at 300 K over the 2*Θ* range of 10-90°, with a step size of 0.01313°, together with the corresponding Rietveld refinement performed using the FullProf suite. All the observed reflections could be indexed using an orthorhombic structure with space group $P2_12_12_1$ (No. 19), consistent with the previously reported crystal structure of $Gd_2TeO_6$ [25]. No additional reflections arising from secondary phases are observed, supporting the formation of a single-phase sample. The refinement yields lattice parameters of $a$ = 5.3194(2) Å, $b$ = 9.1524(3) Å, and $c$ = 10.0485(4) Å, with a unit-cell volume of $V$ = 489.21(3) Å$^3$. The quality of the refinement is evidenced by $R_p$ = 2.67 %, $R_{wp}$ = 5.60 %, $R_{exp}$ = 3.09 %, and $\chi^2$ = 3.29.

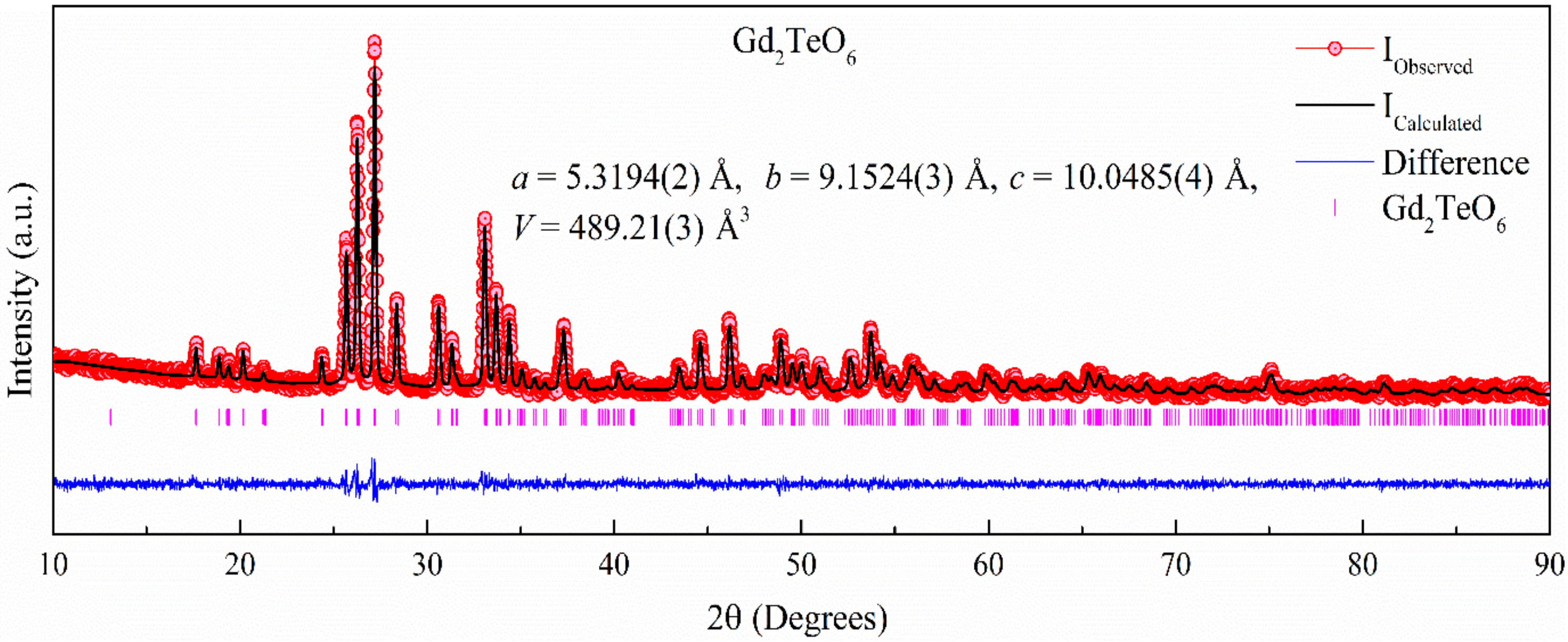

**Fig.1.** Experimentally observed PXRD pattern of $Gd_2TeO_6$ along with the Rietveld refinement**.**

The refined crystal structure of $Gd_2TeO_6$ is presented in **Fig. 2(a)**. The structure contains two crystallographically independent Gd sites (Gd1 and Gd2), one Te site, and six crystallographically distinct oxygen sites. Both Gd sites exhibit seven-fold coordination with oxygen, although their local coordination geometries are different: Gd1 forms a square-triangular $GdO_7$ polyhedron, whereas Gd2 adopts a capped trigonal-prismatic $GdO_7$ environment. The $Gd1O_7$ and $Gd2O_7$ polyhedra are connected along the [100] direction to form zigzag chains, through edge sharing for the Gd1 polyhedra and corner sharing for the Gd2 polyhedra. These chains further interconnect to generate a three-dimensional Gd-O framework. In contrast, the $Te^{6+}$ cation is coordinated by six oxygen atoms, forming a slightly distorted $TeO_6$ octahedron. The $TeO_6$ octahedra remain isolated from one another and are accommodated within the channels of the Gd-O framework. The corresponding local coordination environments and Gd-O and Te-O bond distances are shown in **Fig. 2(b-d)**, respectively. These structural features are consistent with the previously reported single-crystal structure of $Gd_2TeO_6$ [25]. Furthermore, the atomic coordinates obtained from the Rietveld refinement are shown in the **Table -SI** of the Supplementary Information section.

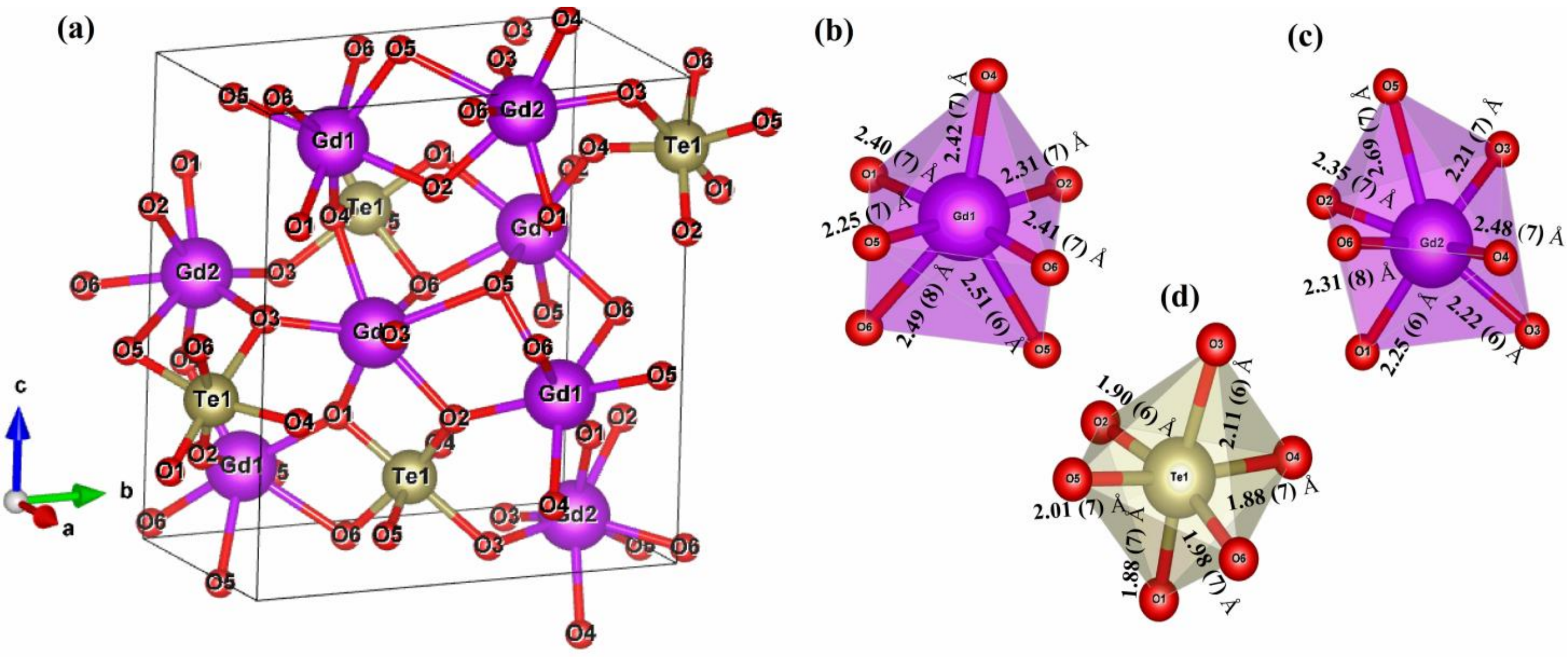

**Fig. 2.** Crystal structure and local coordination environments of $Gd_2TeO_6$: **(a)** crystal structure, **(b)** square-triangle $Gd1O_7$, **(c)** capped trigonal-prismatic $Gd2O_7$, and **(d)** distorted $TeO_6$ octahedron.

### 3.2.Temperature-dependent magnetic properties: *M* (T)

To investigate the magnetic properties of $Gd_2TeO_6$, temperature-dependent magnetization measurements, *M*(T), were performed in the temperature range of 2-300 K under applied magnetic fields of 100 Oe, 5 kOe, and 5 T. Main panel of **Fig.3. (a)** shows the *M*(T) curves measured under an applied field of 100 Oe. The magnetization increases progressively upon cooling, with a pronounced upturn in the low-temperature region, indicative of a predominantly paramagnetic response arising from the $Gd^{3+}$ magnetic moments. No sharp anomaly characteristic of a conventional long-range magnetic transition is observed down to the lowest measured temperature of 2 K. A magnified view of the low-temperature region, however, reveals a distinct bifurcation between the field-cooled cooling (FCC) and field-cooled warming (FCW) branches. A similar bifurcation is also observed under an applied field of 5 kOe, with the onset of the separation initiating from approximately 3.7 K for both fields. Conversely, the FCC, FCW, and ZFC curves become effectively coincident under the applied field of 5 T, as can be seen in the upper inset of **Fig. 3. (a)**. The field dependence of this low-temperature irreversibility indicates that the observed feature is sensitive to the applied magnetic field and is associated with low-energy magnetic correlations or slow magnetic processes. Such behaviour can originate from several mechanisms, including slow magnetic relaxation, spin freezing, or domain wall-related effects [30,31]. However, the present DC magnetization measurements alone are limited to uniquely distinguish between these possibilities, and frequency-dependent AC susceptibility or magnetic relaxation measurements would be essential for a definitive assignment.

To further analyze the low-temperature magnetic feature, the first derivative of the susceptibility, $d\chi/dT$ was calculated from the 100 Oe data as shown in the lower inset of **Fig. 3. (a)**. The $d\chi_{ZFC}/dT$, and $d\chi_{FCW}/dT$ curves exhibit a similar feature with a minimum around 2.6 K, whereas the corresponding FCC branch does not show a pronounced anomaly and remains nearly featureless down to the lowest measured temperature. The characteristic temperature obtained from the derivative analysis is close to the onset of the low-field bifurcation observed in the magnetization curves, indicating a change in the low-temperature magnetic response around this temperature. However, since no sharp anomaly associated with a conventional long-range magnetic transition is observed, this feature is more appropriately regarded as a low-temperature magnetic anomaly/irreversibility rather than definitive evidence for a spin-glass or spin-frozen state.

The nature and strength of the magnetic interactions were further investigated using a Curie-Weiss analysis of the inverse susceptibility obtained from the 5 kOe ZFC data. The inverse susceptibility shows a linear temperature dependence in the high-temperature region, and the data were fitted between 50 and 300 K using the Curie-Weiss law,

$$\chi = \frac{C}{T-\theta_{CW}} \quad (1)$$

where, $C$ is the Curie constant, and $\theta_{CW}$ is the Curie-Weiss temperature, and the resulting fit is shown in **Fig. 3. (b)**. The fitting yields $\theta_{CW}$ = -1.25 K, indicating weak predominant antiferromagnetic interactions and a relatively low characteristic magnetic interaction energy scale. Importantly, the small magnitude of $\theta_{CW}$ is considerably lower than that reported for the structurally related $Gd_2MoO_6$ and $Gd_2WO_6$ compounds, for which $\theta_{CW}$ values of approximately -6.4 K have been reported [29,32]. In $Gd_2MoO_6$, the negative Curie-Weiss temperature is associated with predominant antiferromagnetic interactions, while its specific-heat measurements reveal magnetic ordering at approximately 0.98 K and 0.60 K [33]. The reported analysis of $Gd_2MoO_6$ further indicates that the exchange scale estimated from the Curie-Weiss temperature can differ from that obtained from low-temperature specific heat, highlighting the possible contribution of additional interactions such as crystal-field effects when the exchange energy scale is small. For $Gd_2WO_6$, the reported $\theta_{CW}$ of -6.4 K likewise indicates predominant antiferromagnetic interactions, although the absence of a clear ordering anomaly above 2 K. Therefore, the substantially smaller $\theta_{CW}$ observed for $Gd_2TeO_6$ suggests a weaker overall magnetic interaction scale compared with these related Gd-based oxides. Since no conventional magnetic transition is detected down to 2 K in the present measurements, the possibility of magnetic ordering at a temperature below the experimental limit cannot be excluded. The reduced interaction scale may also favour the retention of magnetic entropy at cryogenic temperatures, which is relevant to the large magnetocaloric response observed in $Gd_2TeO_6$.

The Curie constant obtained from the Curie-Weiss fitting gives an effective magnetic moment of $\mu_{eff}$ of 11.09 $\mu_B$/f.u.. For $Gd_2TeO_6$, the magnetic response is expected to originate predominantly from the two $Gd^{3+}$ ions, each having the $4f^7$ electronic configuration with ($S$ = 7/2), ($L$=0), and ($J$ =7/2). Using ($g_J$ = 2), the theoretical effective moment for an individual $Gd^{3+}$ ion is $\mu_{eff} = g_J\sqrt{J(J+1)}$, for the $Gd^{3+}$ ion, therefore ≈ 7.94$\mu_B$. Consequently, for two magnetically independent $Gd^{3+}$ ions per formula unit, the expected effective moment is $\sqrt{2}$ ≈ 11.23 $\mu_B$/f.u.. which is in close agreement with the experimentally obtained value of 11.09 $\mu_B$/f.u.. This agreement confirms that the magnetic susceptibility is predominantly governed by the localized $Gd^{3+}$ ($4f^7$) moments. The combination of the large spin moment of $Gd^{3+}$ and the relatively weak net magnetic interaction inferred from the small $\theta_{CW}$ is therefore particularly relevant to the preservation and field manipulation of magnetic entropy at low temperatures.

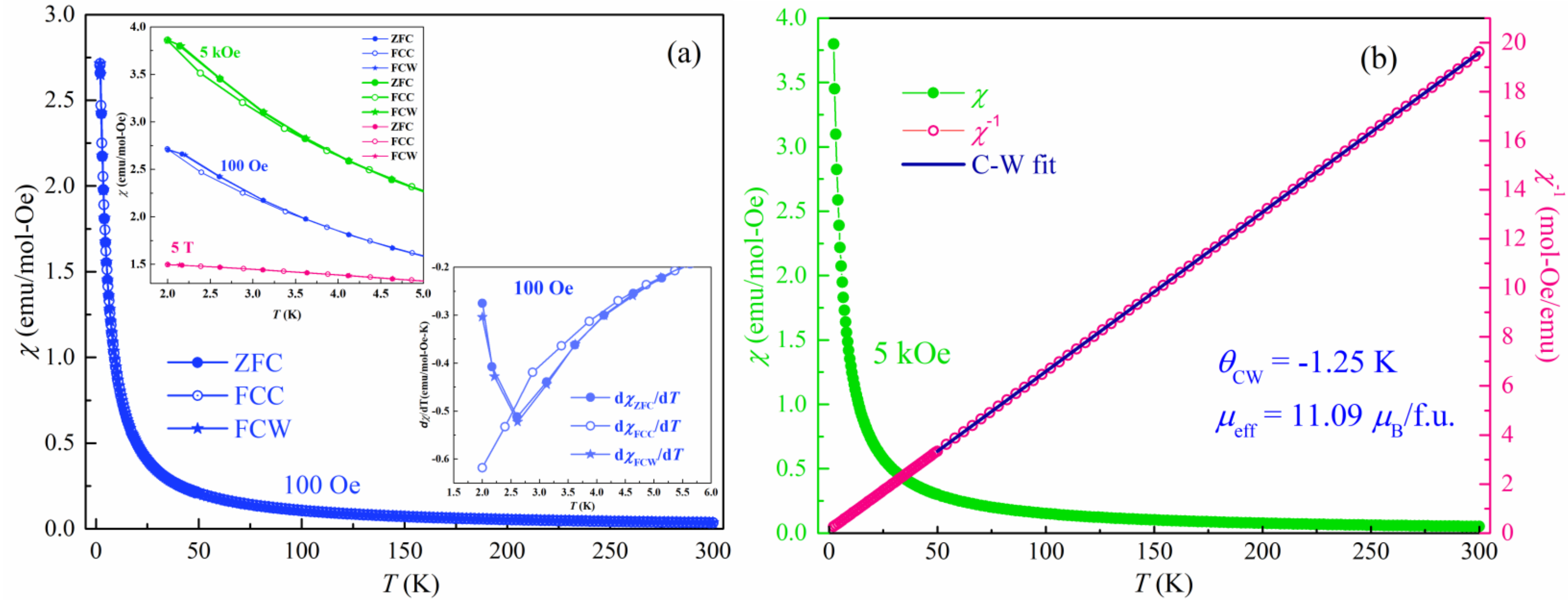

**Fig. 3.** **(a)** Temperature-dependent susceptibility of $Gd_2TeO_6$ in ZFC, FCC, and FCW protocol with a field of 100 Oe, left inset showing the enlarged view of magnetization with different fields, whereas, the right-bottom inset shows the enlarged view of d$\chi$/d$T$ for different protocols with a field of 100 Oe, and **(b)** Temperature dependence of the susceptibility along with the inverse of it for the field of 5 kOe along with the Curie-Weiss fit.

### 3.3. Field dependent magnetization: *M* (H)

To gain further insight into the magnetic response and field-induced polarization of $Gd_2TeO_6$, isothermal magnetization measurements, *M*(H), were performed at different temperatures in magnetic fields up to 7 T, as shown in **Fig. 4. (a)**. At 2 K, the magnetization exhibits a pronounced nonlinear field dependence, increasing rapidly at low fields and gradually approaching saturation at higher fields. The rapid increase is particularly evident below approximately 3 T, followed by a progressive reduction in the field-induced change in magnetization at higher fields. At 7 T, the magnetization reaches approximately 13.6 $\mu_B$/f.u., corresponding to about 97 % of the theoretical saturation value of 14 $\mu_B$/f.u. expected for two $Gd^{3+}$ ions with $S = 7/2$. Thus, the Gd moments are already strongly polarized at 7 T, although measurements at lower temperatures and higher fields would be required to establish the complete saturation limit.

The ascending and descending branches of the 2 K isotherm show complete overlap, indicating negligible magnetic hysteresis within the experimental resolution. The absence of a substantial hysteresis loss is favourable for reversible magnetic-field cycling [34,35]. With increasing temperature, the *M* (H) curves progressively become more linear, reflecting the increasing dominance of the paramagnetic response over the low-temperature magnetic correlations. The corresponding d*M*/d*H* curves (shown in the inset of **Fig. 4. (a)**) further illustrate this evolution. At 2 K, d*M*/d*H* is largest at low fields and decreases progressively with increasing field, consistent with the rapid initial polarization of the Gd moments followed by a gradual approach toward saturation. With increasing temperature, the magnitude of the field response decreases and the field dependence becomes increasingly linear. Above approximately 40 K, the *M* (H) isotherms are nearly linear over the investigated field range (as shown in **Fig. S1. (a)**), consistent with a predominantly paramagnetic regime.

To examine the high-field saturation behaviour, additional *M* (H) measurements were performed at different temperatures. **Fig. 4. (b)** shows the comparison between the magnetization obtained till 7 T and 14 T at 3 K. The magnetization curves obtained using the MPMS and PPMS follow essentially the same trajectory in their common field range up to 7 T, confirming the consistency of the measurements from the two instruments. At 3 K, the magnetization increases from approximately 13.41 $\mu_B$/f.u. at 7 T to 13.74 $\mu_B$/f.u. at 14 T, corresponding to an increase of only about 2.4 %. This relatively small additional increase indicates that the Gd moments are already close to saturation at 7 T, with only a limited further polarization occurring at higher fields. The isothermal field-dependent magnetization, obtained up to 14 T, for different temperatures are shown in **Fig. S1. (b).**

**Fig. 4. (c)** shows the comparison between the theoretical magnetization from the Brillouin function for magnetization at 2 K, $B_J(x)$. The deviation of the experimental *M* (H) curve from the ideal Brillouin response provides further information regarding the magnetic interactions. For comparison, the Brillouin function for non-interacting J = 7/2 spins was calculated using

$$B_J(x) = \frac{2J+1}{2J} coth\left(\frac{2J+1}{2J}x\right) - \frac{1}{2J} coth\left(\frac{x}{2J}\right) \qquad (2)$$

Where the dimensionless parameter, $x = \frac{g_J \mu_B \mu_0 H}{k_B T}$, $g_J = 2$, and $J = 7/2$ were used for $Gd^{3+}$. The calculated magnetization was obtained from $M = N g_J \mu_B J B_J(x)$, considering two $Gd^{3+}$ ions per formula unit. As shown in **Fig. 4. (c)**, the experimental magnetization at 2 K lies systematically below the non-interacting Brillouin prediction, with the maximum difference occurring at approximately 0.8-1 T. At higher fields, the difference decreases and becomes nearly constant above approximately 5 T, as shown in the inset of **Fig. 4. (c)**. Since the Brillouin function describes independent magnetic moments, the observed deviation indicates the presence of inter-ionic magnetic interactions that are not included in the non-interacting model [12,36]. In association with the negative Curie-Weiss temperature $\Theta_{CW}$ = -1.25 K, the observed deviation is consistent with predominant antiferromagnetic correlations in $Gd_2TeO_6$. The stronger deviation at low and intermediate fields can be understood as a result of the competition between the exchange interactions and the applied-field-induced Zeeman energy. As the magnetic field increases, the Zeeman energy progressively overcomes the relatively weak antiferromagnetic interactions and drives the Gd moments toward a polarized state, resulting in a reduced deviation from the ideal paramagnetic saturation behaviour at high fields.

A comparison with the related $Gd_2WO_6$ and $Gd_2MoO_6$ compounds provides further context for the magnetic response of $Gd_2TeO_6$. $Gd_2WO_6$ exhibits a pronounced deviation from the Brillouin function at 2 K and reaches only approximately 6.5 $\mu_B$ per $Gd^{3+}$ at 7 T, which was attributed to substantial antiferromagnetic interactions and incomplete field-induced saturation [29]. The reported Curie-Weiss temperature of $Gd_2WO_6$ is approximately -6.4 K, considerably larger in magnitude than that of $Gd_2TeO_6$. Similarly, $Gd_2MoO_6$ requires fields extending to 9 T to approach its saturation magnetization [33]. These comparisons suggest that the smaller magnitude of $\theta_{CW}$ in $Gd_2TeO_6$ is consistent with a comparatively weaker overall magnetic interaction scale, allowing the Gd moments to become more readily polarized by the applied field. However, $\theta_{CW}$ represents a net interaction scale and does not by itself determine the magnitude of individual exchange parameters; the microscopic origin of the magnetic interactions is therefore examined further through the first-principles calculations.

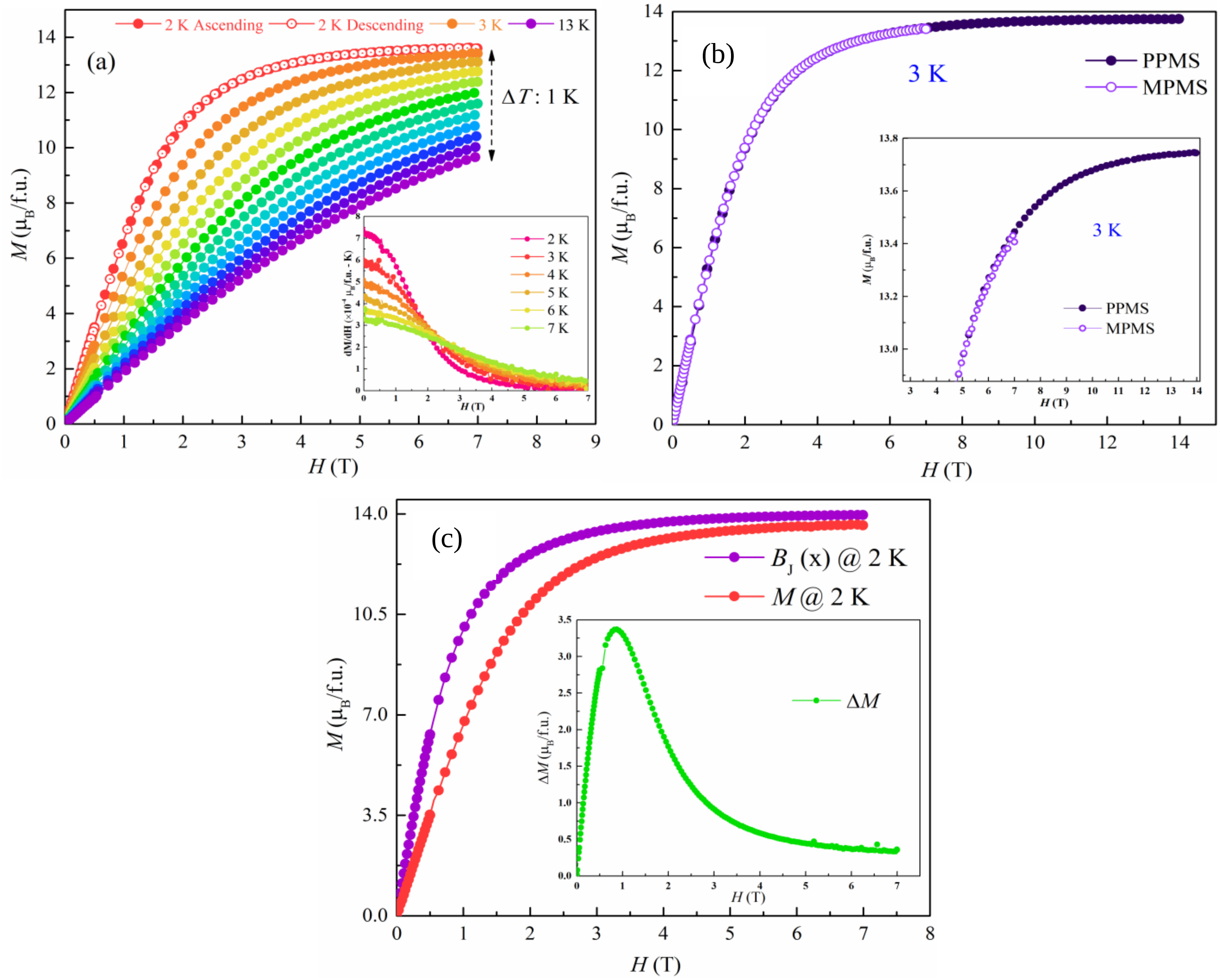

**Fig. 4. (a)** Field-dependent magnetization of Gd2TeO6, at different temperatures upto the field range of 7 T, inset showing the d*M*/d*H* behaviour, **(b)** Comparision between the Field-dependent magnetization obtained at 3 K for the field ranges of 7 T and 14 T, inset showing the enlarged view of the magnetization in the 5 T–14 T region, and **(C)** Comparision between the Brillouin function for magnetization and the experimentally obtained magnetization at 2 K, inset showing the difference between the two magnetizations for 2 K

### 3.4. Density functional theory (DFT) calculations:

To obtain microscopic insight into the electronic and magnetic properties of $Gd_2TeO_6$, Spin-polarized first-principles DFT calculations were performed for ten collinear magnetic configurations, consisting of seven AFM, two ferrimagnetic (FiM), and one ferromagnetic (FM) arrangement, using the crystal structure obtained from the Rietveld refinement of the room-temperature PXRD. The eight Gd atoms in the computational unit cell originate from two crystallographically distinct Gd sites, with four atoms corresponding to each Gd sublattice. The corresponding spin configurations are provided in **Fig. S2**, and **Fig. S3**.

Initially, calculations were performed within the GGA formalism with the electronic SCF convergence criteria of $10^{-6}$ eV, and the obtained energies are given in **Table-SII**. Among the considered configurations, $AF_4$ exhibits the lowest total energy of -334.72376141 eV, followed by $AF_1$ and $AF_3$, with total energies of -334.72354001 eV and -334.72318816 eV, respectively, giving the hierarchy, $E_{\mathrm{AF}_4} < E_{\mathrm{AF}_1} < E_{\mathrm{AF}_3}$. The $AF_4$ - $AF_1$ energy difference is only 0.221 meV per unit cell, indicating competing AFM states. To verify that this small energy separation was not an artefact of the SCF convergence criterion, the GGA calculations were repeated for the closely competing configurations using a tighter SCF convergence criteria of $10^{-8}$ eV. The resulting energy hierarchy remained unchanged, with $AF_4$ still lower in energy than $AF_1$ and $AF_3$, confirming the numerical consistency of the GGA (the total energy obtained are shown in **Table- SIII)**. The $AF_4$ configuration corresponds to the ↑↑↓↓↓↓↑↑ spin structure, in which the four Gd1-derived moments follow a ↑↑↓↓ arrangement and the four Gd2-derived moments follow the ↓↓↑↑ arrangement. The corresponding spin-resolved projected density of states (PDOS) indicates an insulating state with a gap of approximately 2.8 eV, with O-2*p* states dominating near the valence-band edge and localized Gd-4f states separated from the Fermi level, as depicted in **Fig. 5. (a)**.

Because of the localized nature of the Gd-4*f* electrons, the total energies of configurations were subsequently recalculated using GGA + $U$ with $U$ = 6 eV and SCF convergence criteria of $10^{-8}$ eV, and the obtained energies are given in **Table- SIV**. In contrast to GGA, $AF_3$ becomes the lowest-energy configuration, with a total energy of -333.69862384 eV, followed by $AF_1$ with -333.69806346 eV; AF3 is therefore lower by 0.560 meV per unit cell, while $AF_4$ lies 1.285 meV above $AF_3$. The $AF_3$ configuration, represented by ↑↑↑↑↓↓↓↓, consists of four mutually parallel Gd1-derived moments and four mutually parallel Gd2-derived moments, with the two groups oriented antiparallel to each other. Thus, inclusion of $U$ changes the preferred arrangement from the internally modulated $AF_4$ state to an inter-sub lattice AFM arrangement.

The robustness of the $AF_3$-$AF_1$ energy difference was further examined using a denser 8*6*5 k-point mesh and $U$ = 5-8 eV. AF3 remained lower in energy than AF1, with differences of 0.566, 0.606, 0.613, and 0.582 meV per unit cell for 5, 6, 7, and 8 eV, respectively, indicating that the relative stability of $AF_3$ is not significantly affected by the $U$ values or k-point mesh sampling (the calculated energy values are given in **Table- SV**). **Fig. 5. (b)** shows the PDOS of $Gd_2TeO_6$ with $U$ = 6 eV for $AF_3$ configuration, indicating an insulating gap of approximately 3 eV, with the Gd-4*f* states shifted further away from the Fermi level, consistent with their localized character. The calculated local moments of Gd1 and Gd2 are 7.044 $\mu_B$ and 7.055 $\mu_B$, for $U = 6$ eV respectively, which are close to the expected spin moment of $Gd^{3+}$.

Overall, the calculations reveal closely competing AFM configurations in $Gd_2TeO_6$. While GGA favors AF4, inclusion of the Hubbard correction stabilizes AF3, which remains robust against variations in $U$ and k-point sampling. The presence of several low-lying AFM states indicates a relatively narrow magnetic-energy landscape, which may be relevant to the complex low-temperature magnetic response observed experimentally.

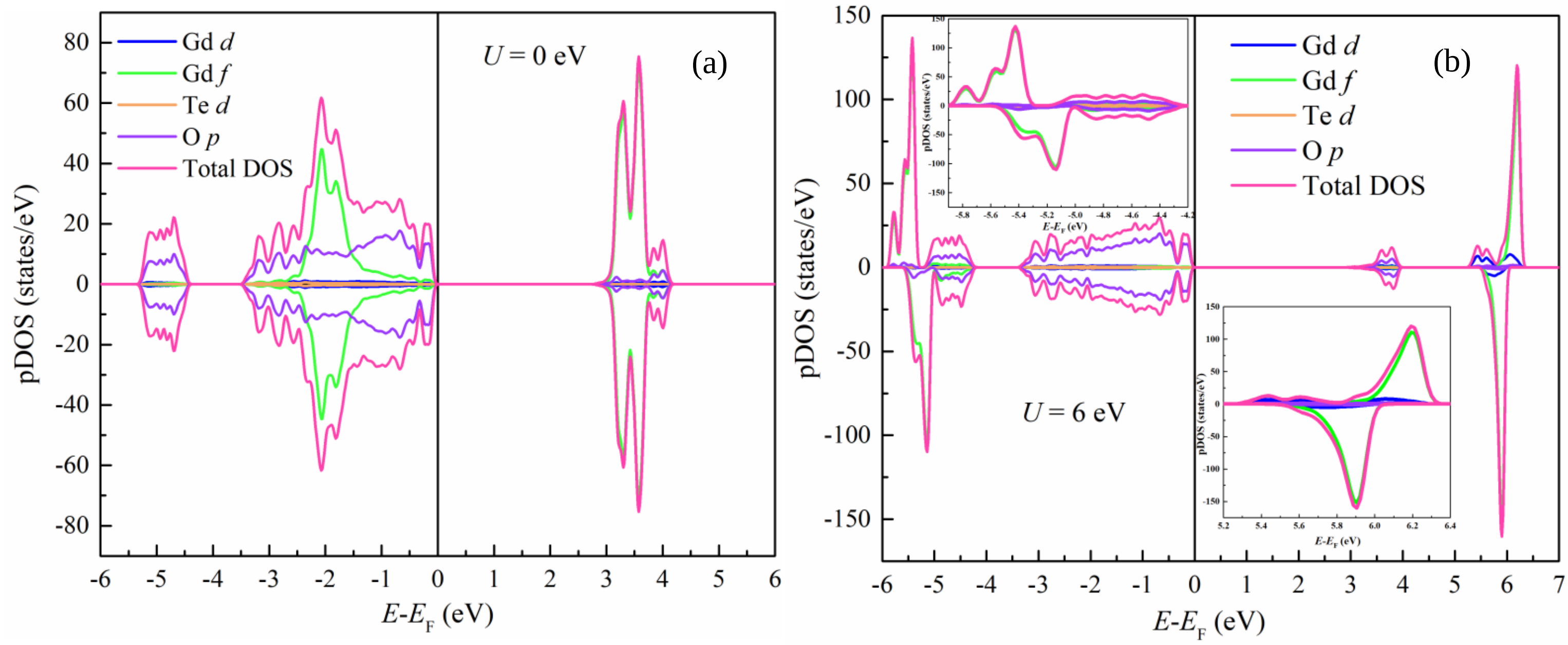

**Fig. 5.** Partial density of states of selected orbitals of constituent atoms as obtained from **(a)** GGA formalism and **(b)** GGA+ $U$ ($U = 6$ eV), with the fermi energy ($E_F$) set to 0 eV.

### 3.5. Specific heat capacity ($C_P$):

To further investigate the low-temperature magnetic behaviour and thermodynamic response of $Gd_2TeO_6$, heat-capacity measurements were performed between 2 and 250 K in zero magnetic field and between 2 and 60 K under applied magnetic fields of 5, 7, and 14 T, as shown in **Fig. 6. (a)**. In a zero magnetic field, no sharp $\lambda$-type anomaly is observed over the measured temperature range, indicating the absence of a long-range magnetic ordering transition above 2 K. This observation is consistent with the absence of a clear magnetic transition in the dc magnetization measurements. Nevertheless, the zero-field $C_P$(T) curve exhibits a pronounced upturn below approximately 8 K, suggesting the presence of substantial low-energy magnetic excitations or correlations in this temperature regime.

Upon application of an external magnetic field, the low-temperature heat capacity is strongly modified. A broad maximum develops in the $C_P$(T) curves and progressively shifts toward higher temperature with increasing magnetic field as shown in **Fig. 6. (a)**. Such a field-dependent broad maximum is characteristic of a field-induced redistribution of magnetic entropy and can arise from the Zeeman splitting of the degenerate $4f^7$ ground-state multiplets of $Gd^{3+}$. Similar field-induced broad maxima have been reported in Gd based magnetic materials and have been associated with the evolution of low-energy magnetic excitations and Schottky-type contributions arising from field induced level splitting [37,38].

The absence of a sharp zero field anomaly, together with the negative Curie-Weiss temperature obtained from the magnetic susceptibility, suggests that the low-temperature magnetic response is dominated by weak antiferromagnetic interactions and short-range magnetic correlations rather than a well-established long-range ordered state above 2 K. However, the possibility of magnetic ordering at temperatures below the present experimental limit cannot be excluded. Measurements extending below 2 K would therefore be required to determine whether a thermodynamic transition occurs in this temperature region.

At higher temperatures, the $C_P$(T) curves measured at 0, 5, and 7 T become increasingly similar, particularly above approximately 50 K, whereas the 14 T curve remains noticeably enhanced. This indicates that the influence of the applied magnetic field extends to higher temperatures for 14 T, consistent with a field dependent redistribution of magnetic entropy associated with the $Gd^{3+}$ magnetic levels.

For an insulating compound such as $Gd_2TeO_6$, the electronic contribution to the heat capacity is expected to be negligible, and the total heat capacity can therefore be expressed approximately as

$$C_P(T) = C_{Ph}(T) + C_M(T) \qquad (3)$$

Where, $C_{Ph}$(T) and $C_M$(T) represent the lattice and magnetic contributions, respectivel. The lattice contribution contains contributions from both acoustic and optical phonon modes. Accordingly, a Debye-Einstein model incorporating one Debye term and two Einstein terms was employed to describe the phonon contribution given by,

$$C_{phonon} = (1-p-q)9nR\left(\frac{T}{\Theta_D}\right)^3 \int_0^{\Theta_D/T} \frac{x^4 e^x}{(e^x-1)^2}dx + 3nR\left[p\,\frac{\left(\frac{E_1}{x}\right)^2 exp\left(\frac{E_1}{x}\right)}{\left(exp\left(\frac{E_1}{x}\right)-1\right)^2} + q\,\frac{\left(\frac{E_2}{x}\right)^2 exp\left(\frac{E_2}{x}\right)}{\left(exp\left(\frac{E_2}{x}\right)-1\right)^2}\right] \qquad (4)$$

Where, $n$ is the number of atoms, which was set to be 9 for $Gd_2TeO_6$, $R$ is the universal gas constant, $(1-p-q)$ is the fraction of vibrational modes contributing to the Einstein-type heat capacity, $\Theta_D$ is the Debye temperature and $E_1$, and $E_2$ are the Einstein temperatures [39,40]. The zero-field heat-capacity data in the range 15-230 K were fitted using this model. The fit yields a $\Theta_D$ of 398(11) K and Einstein temperatures of $E_1$ and $E_2$ to be 117(23) K and 814(10) K respectively. The fitted phonon contribution reproduces the experimental heat capacity well over the fitting range and was further extrapolated toward lower temperatures to estimate the lattice contribution in the region where the magnetic contribution becomes significant. The corresponding fitted phonon background is shown together with the experimental zero-field $C_P$(T) data in **Fig. 6. (b)**.

The magnetic contribution was then obtained according to $C_M$(T) = $C_P$(T) - $C_{Ph}$(T). The resulting $C_M$(T)/T curves for different magnetic fields are presented in **Fig. 6. (c)**. At zero field, $C_M$(T)/T increases continuously toward the lowest measured temperature without exhibiting a well-defined sharp anomaly above 2 K. With increasing magnetic field, a broad maximum develops and shifts progressively toward higher temperatures. The evolution of $C_M$(T)/T closely follows the corresponding field dependence observed in $C_P$(T), confirming the predominantly magnetic origin of the low temperature field-dependent contribution. The broadening and systematic temperature shift of this feature are consistent with field-induced splitting and redistribution of the $Gd^{3+}$ magnetic levels rather than with a conventional long-range ordering transition.

The magnetic entropy was subsequently estimated from $S_M(T) = \int_{T_{min}}^{T} \frac{C_M}{T}$ ,where $T_{\min}$ = 2 K is the lowest experimental temperature. The resulting $S_M(T)$ curves are shown in the inset of **Fig. 6. (c)**. The zero field entropy reaches only approximately 3 J mol$^{-1}$ K$^{-1}$ within the measured temperature interval. This value is considerably smaller than the expected magnetic entropy associated with two $Gd^{3+}$ ions per formula unit, $s_M^{max} = 2Rln(2J+1) = 2Rln8$ = 34.57 J mol$^{-1}$ K$^{-1}$. The low entropy recovered at zero field within the experimentally accessible range indicates a substantial fraction of the magnetic entropy may therefore be associated with the low-temperature region below the measurement limit, a similar phenomena is observed in case of the recently reported heat capacity studies on $KbaGd(BO_3)_2$ and $LnTaO_4$, where, Ln = Tb, Dy, Ho, Er [41,42].

With increasing magnetic field, the magnetic entropy is progressively shifted toward higher values. The $S_M(T)$ curves for 5 T and 7 T approach approximately 29 and 32 J mol$^{-1}$ K$^{-1}$, respectively, within the measured temperature range, while the 14 T curve continues to increase at 60 K and approaches the theoretical $2Rln8$ limit. This field-induced redistribution of magnetic entropy is consistent with the broad field-dependent heat-capacity features and provides direct thermodynamic evidence for the large magnetic entropy available in $Gd_2TeO_6$ at low temperatures.

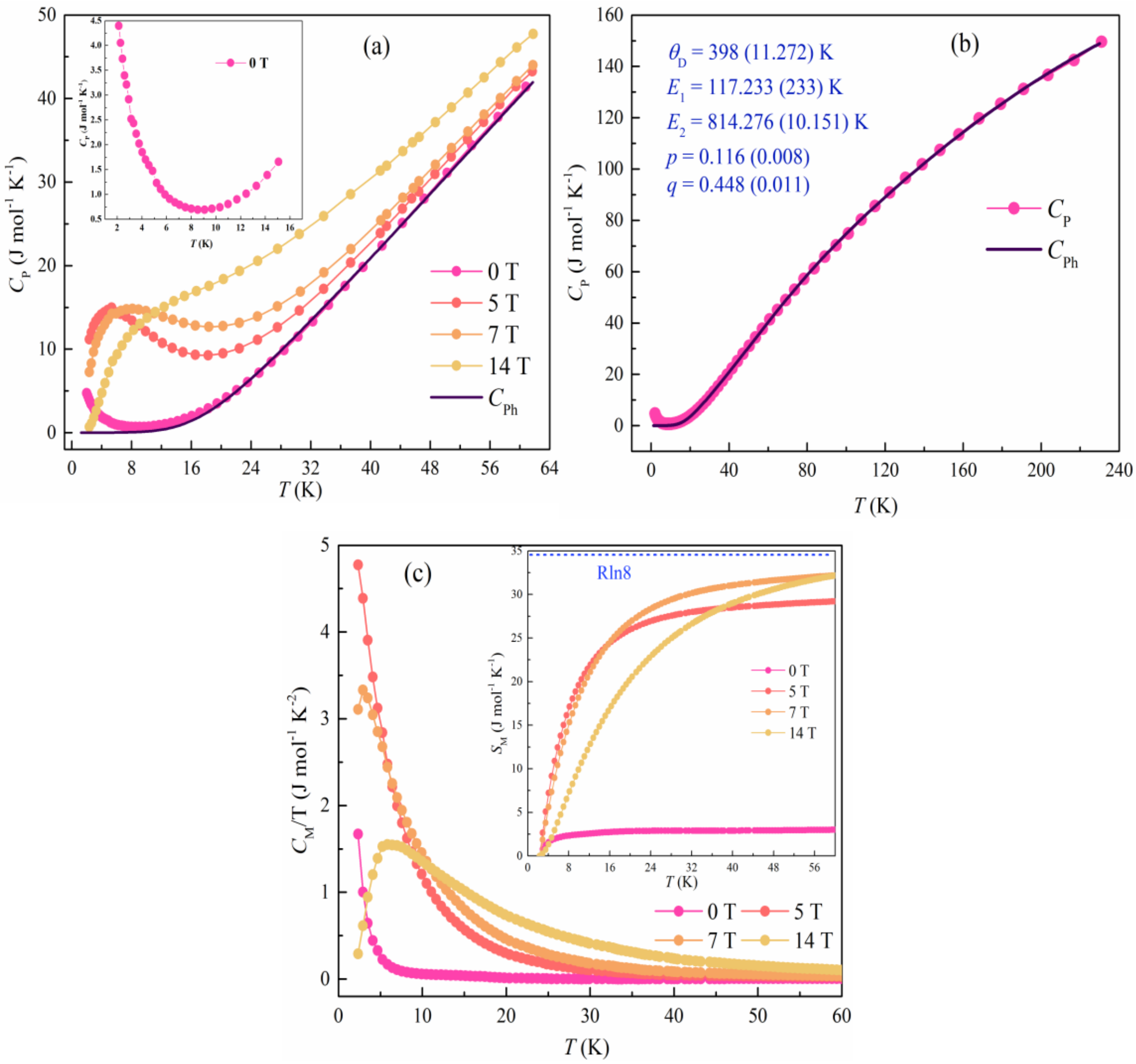

**Fig. 6. (a)** Temperature-dependent specific heat capacity of $Gd_2TeO_6$ with different fields along with the phonon-heat-capacity fit in the temperature range of 2 K – 60 K, inset showing the enlarged view of the heat-capacity with 0 T, **(b)** Temperature-dependent specific heat capacity with field of 0 T in 2 K-230 K interval along with Debye-Einstein model fit for the phonon heat capacity, and **(c)** Temperature dependence of the $C_M$(T)/T with different fields, inset showing the magnetic entropy for different fields.

### 3.6. Magnetocaloric properties:

To evaluate the magnetocaloric response of $Gd_2TeO_6$, the magnetic entropy change $\Delta S_M$, volumetric entropy change $\Delta S_V$, and adiabatic temperature change $\Delta T_{ad}$ were determined from the magnetization and heat-capacity measurements. These parameters provide complementary measures of the refrigerant performance, with $\Delta S_M$ representing the field-induced magnetic entropy change, $\Delta S_V$ accounting for the cooling capacity per unit volume, and $\Delta T_{ad}$ providing an estimate of the temperature change attainable under adiabatic magnetization. The magnetic entropy change was calculated from the isothermal magnetization $M$ (H) curves using the Maxwell thermodynamic relation,

$$\Delta S_M = S(T,H) - S(T,0) = \int_0^H \left(\frac{dM}{dT}\right) dH \tag{5}$$

Since the magnetization measurements were performed at discrete temperatures and fields, the entropy change was evaluated using

$$-\Delta S_M = \sum_i \frac{M_i - M_{i+1}}{T_{i+1} - T_i} \Delta H_i \tag{6}$$

Where $M_i$ and $M_{i+1}$ represent the magnetization values at two consecutive temperatures $T_i$ and $T_{i+1}$, respectively, for a given field interval $\Delta H$ and $T_{avg} = (T_i + T_{i+1})/2$ [43]. The negative sign of $\Delta S_M$ for the present system corresponds to the conventional positive magnetocaloric effect; therefore, the magnitude -$\Delta S_M$is discussed throughout this section.

The temperature dependence of $\Delta S_M$ for different magnetic-field changes is shown in **Fig. 7. (a)**. A pronounced magnetocaloric response is observed over the investigated temperature range, with $\Delta S_M$ increasing progressively with decreasing temperature. At a given temperature, the magnitude of the entropy change increases systematically with increasing magnetic-field change. For field changes of 0-2, 0-5, and 0-7 T, the maximum measured values of $\Delta S_M$ reach 21.40, 48.51, and 54.81 J kg$^{-1}$K$^{-1}$, respectively. The maximum values occur at the lowest measured temperature of 2.5 K, and the curves do not exhibit a peak within the measured temperature range.

The large entropy change at relatively low magnetic fields is particularly significant. At a field change of only 2 T, $Gd_2TeO_6$ exhibits a $\Delta S_M$ of 21.40 J kg$^{-1}$K$^{-1}$ at 2.5 K, while the entropy change reaches 54.81 J kg$^{-1}$K$^{-1}$ for a 0-7 T field change. The magnetization isotherms measured up to 14 T were further used to evaluate the high-field magnetocaloric response. For a field change of 0-14 T, the maximum measured $\Delta S_M$ reaches 63.51 J kg$^{-1}$K$^{-1}$. This value is remarkably close to the theoretical magnetic entropy expected for two $Gd^{3+}$ ions per formula unit, $S_M^{max} = 2Rln(2J+1) = 2Rln8$ = 34.57 J mol$^{-1}$ K$^{-1}$which corresponds to 64.25 J kg$^{-1}$ K$^{-1}$ for $Gd_2TeO_6$.

Thus, the 14 T entropy change corresponds to approximately 98.8 % of the theoretical magnetic entropy limit. The large entropy recovery is consistent with the magnetic nature of the $Gd^{3+}$ sublattice. $Gd^{3+}$ possesses a $4f^7$ electronic configuration with $J = 7/2$, providing an eight-fold magnetic degeneracy. The large fraction of the theoretical entropy accessed under an applied magnetic field indicates that a substantial portion of these magnetic degrees of freedom can be polarized in $Gd_2TeO_6$. The field evolution of the magnetic entropy change provides further insight into the magnetocaloric response. As shown in **Fig. 7. (a)**, $\Delta S_M$ increases rapidly with magnetic field up to approximately 5 T, after which the rate of increase becomes progressively smaller. Above approximately 7 T, only a relatively small additional increase in $\Delta S_M$ is observed. This behavior closely follows the field evolution of the magnetization, where the *M*(H) curves exhibit a rapid increase at low and intermediate fields followed by a progressively weaker field dependence as the $Gd^{3+}$ moments approach saturation.

The rapid increase in magnetization at low and intermediate fields indicates substantial field-induced polarization of the magnetic moments. Upon further increasing the field, the progressive approach toward magnetic saturation reduces the incremental magnetization response and, consequently, the additional magnetic entropy that can be removed by increasing the field. This provides a physical explanation for the experimentally observed rapid increase in $\Delta S_M$ at low-to-intermediate fields followed by a reduced rate of increase at higher fields. It should be noted that the magnetic entropy change is determined thermodynamically by $dM/dT$ through the Maxwell relation rather than directly by $dM/dH$. Therefore, the *M* (H) curves provide the field-polarization context for the observed field dependence of $\Delta S_M$, rather than a direct quantitative measure of the entropy change. The magnetic entropy change was independently evaluated from the heat-capacity measurements performed under zero and finite magnetic fields. The field-induced entropy change can be obtained from

$$\Delta S_M = \int_{T_{min}}^{T} \frac{C(T,H_1)-C(T,H_0)}{T}\,dT \qquad (7)$$

Where, $H_1$ represents the applied magnetic field and $T_{min}$ = 2.36 K is the lowest experimentally measured temperature. The entropy changes obtained from the heat-capacity data are compared with those calculated from the magnetization measurements in **Fig. 7. (a)**. The entropy curves obtained for 5 and 7 T show good agreement with the corresponding magnetization-derived values, providing an independent consistency check of the magnetocaloric analysis. In contrast, the entropy obtained from the 14 T heat-capacity data shows a noticeable deviation from the magnetization-derived 14 T result. The difference at 14 T can be understood in terms of the field dependence of the heat capacity. As discussed above, the 14 T $C_P$(T) curve remains considerably higher than the corresponding 0, 5 and 7 T curves over an extended temperature range, indicating a significant field-dependent contribution to the heat capacity. Moreover, the magnetic entropy $\Delta S_M$ obtained by integrating $C_M$(T)/T at 14 T continues to increase up to the highest measured temperature of 60 K as shown in **Fig. 6. (c)**. Consequently, the heat-capacity integration does not fully capture the complete entropy recovery within the available temperature range. High-temperature heat-capacity measurements under 14 T would therefore be needed to establish the complete entropy change from this approach. The reasonable agreement between the magnetization, and heat-capacity derived entropy changes at 5 and 7 T nevertheless supports the reliability of the magnetocaloric analysis.

For practical magnetic refrigeration, the volumetric entropy change is an important parameter because it determines the cooling capacity per unit volume of the refrigerant and is hence directly relevant to the compactness of a refrigeration device [16]. The volumetric magnetic entropy change was calculated according to

$$\Delta S_V = \rho \Delta S_M \qquad (8)$$

Where, $\rho$ is the density of $Gd_2TeO_6$, where, the volume of the cell obtained from the PXRD refinement was considered. The temperature dependence of the volumetric entropy change is presented in **Fig. 7. (b)**. For field changes of 0-2, 0-5 and 0-7 T, the maximum values of $\Delta S_V$ are 156.31, 354.47 and 400.50 mJ cm$^{-3}$ K$^{-1}$, respectively. For the 0-14 T field change, the maximum volumetric entropy change reaches 464.03 mJ cm$^{-3}$ K$^{-1}$ at approximately 2.75 K. The systematic increase in $\Delta S_V$ with magnetic-field change follows the corresponding enhancement of the gravimetric magnetic entropy change. The relatively large volumetric entropy change observed for $Gd_2TeO_6$, particularly the value of 400.50 mJ cm$^{-3}$ K$^{-1}$ at 7 T, indicates that the material combines a substantial mass-specific magnetocaloric response with a high entropy change per unit volume. This characteristic is relevant for the development of compact cryogenic magnetic refrigeration systems.

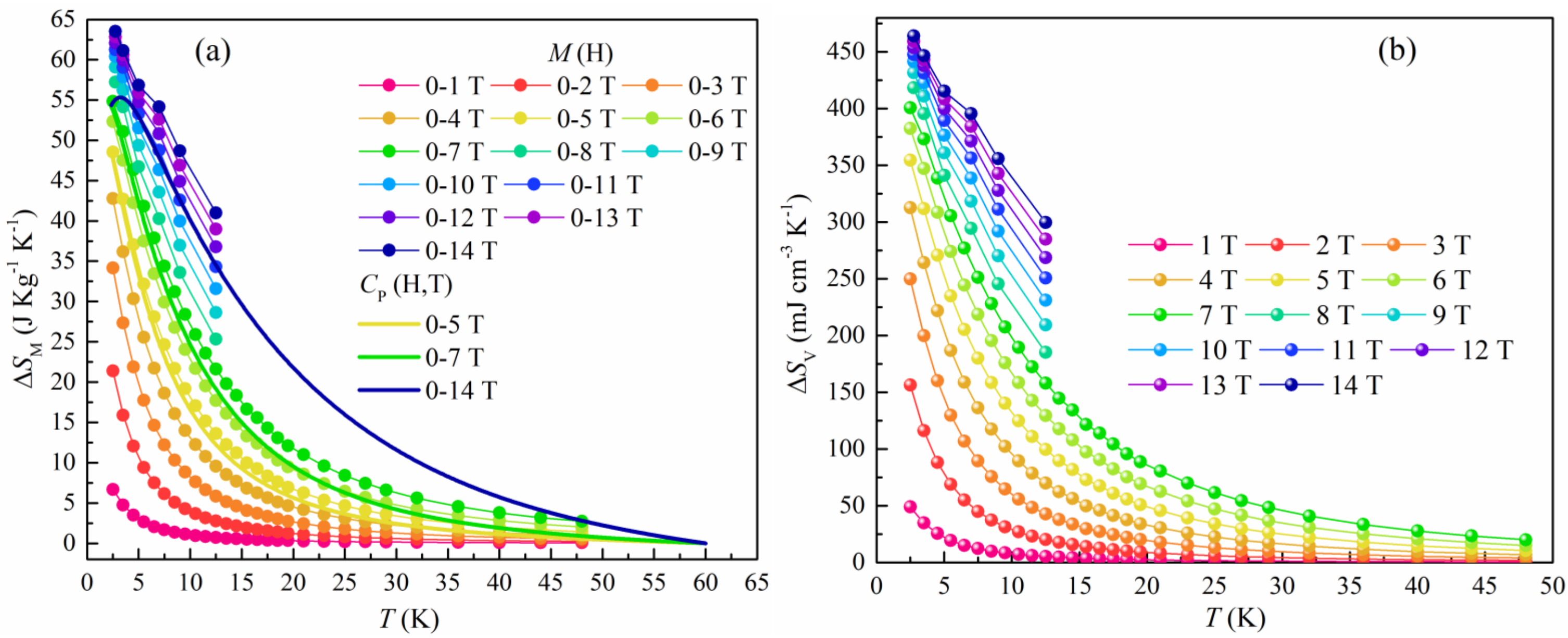

**Fig. 7. (a)** Temperature-dependent gravimetric magnetic entropy change ($\Delta S_M$) of $Gd_2TeO_6$ for different fields, line + symbol and solid lines indicating the $\Delta S_M$ obtained from field-dependent magnetization and the heat-capacity, respectively. **(b)** Temperature-dependent volumetric magnetic entropy change ($\Delta S_V$) of $Gd_2TeO_6$ for different fields.

While the magnetic entropy change provides a measure of the entropy transferred during a magnetic-field change, the adiabatic temperature change $\Delta T_{ad}$ provides a direct indication of the temperature change that can potentially be generated when the magnetic field is varied under adiabatic conditions. Therefore, $\Delta T_{ad}$ represents an important parameter for evaluating the magnetocaloric performance of $Gd_2TeO_6$. The indirect $\Delta T_{ad}$ was estimated using the measured heat-capacity and magnetization data according to

$$\Delta T_{ad} = -\int_{H_i}^{H_f} \left(\frac{T}{C_H(T,H)}\right)\left(\frac{dM}{dT}\right)_H dH \qquad (9)$$

Where, $C_H(T,H)$ is the heat capacity at the corresponding magnetic field, and $H_i$ is the initial field and $H_f$ is the final field corresponsing to $C_H$ [41]. The calculated temperature-dependent $\Delta T_{ad}$ values for field changes of 0-2, 0-5, 0-7 and 0-14 T are shown in **Fig. 8.**. For field changes of 0-2, 0-5 and 0-7 T, the calculated maximum $\Delta T_{ad}$ values are approximately 7.8, 18.3 and 22.7 K, respectively. The maximum for the 0-7 T field change occurs in the vicinity of 12 K. The calculated temperature change remains appreciable even at the lowest measured temperature. At 2.5 K, the calculated $\Delta T_{ad}$ values for field changes of 0-2, 0-5 and 0-7 T are approximately 3.5, 6.6 and 7.5 K, respectively. The presence of a sizable calculated temperature change at the lowest measured temperature is relevant to the cryogenic application of $Gd_2TeO_6$, although measurements extending to lower temperatures would be required to directly establish its behavior in the sub-kelvin regime. For the 0-14 T field change, the resulting $\Delta T_{ad}$ values are approximately 11.99, 16.14, 22.29, 27.00 and 31.74 K at 3.5, 5, 7, 9 and 12.5 K, respectively. It should be noted that the values obtained here represent an indirect integration-based estimate derived from magnetization and heat-capacity measurements rather than a direct adiabatic measurement. Nevertheless, the combination of a large entropy change and a substantial calculated temperature change demonstrates the strong magnetocaloric response of $Gd_2TeO_6$.

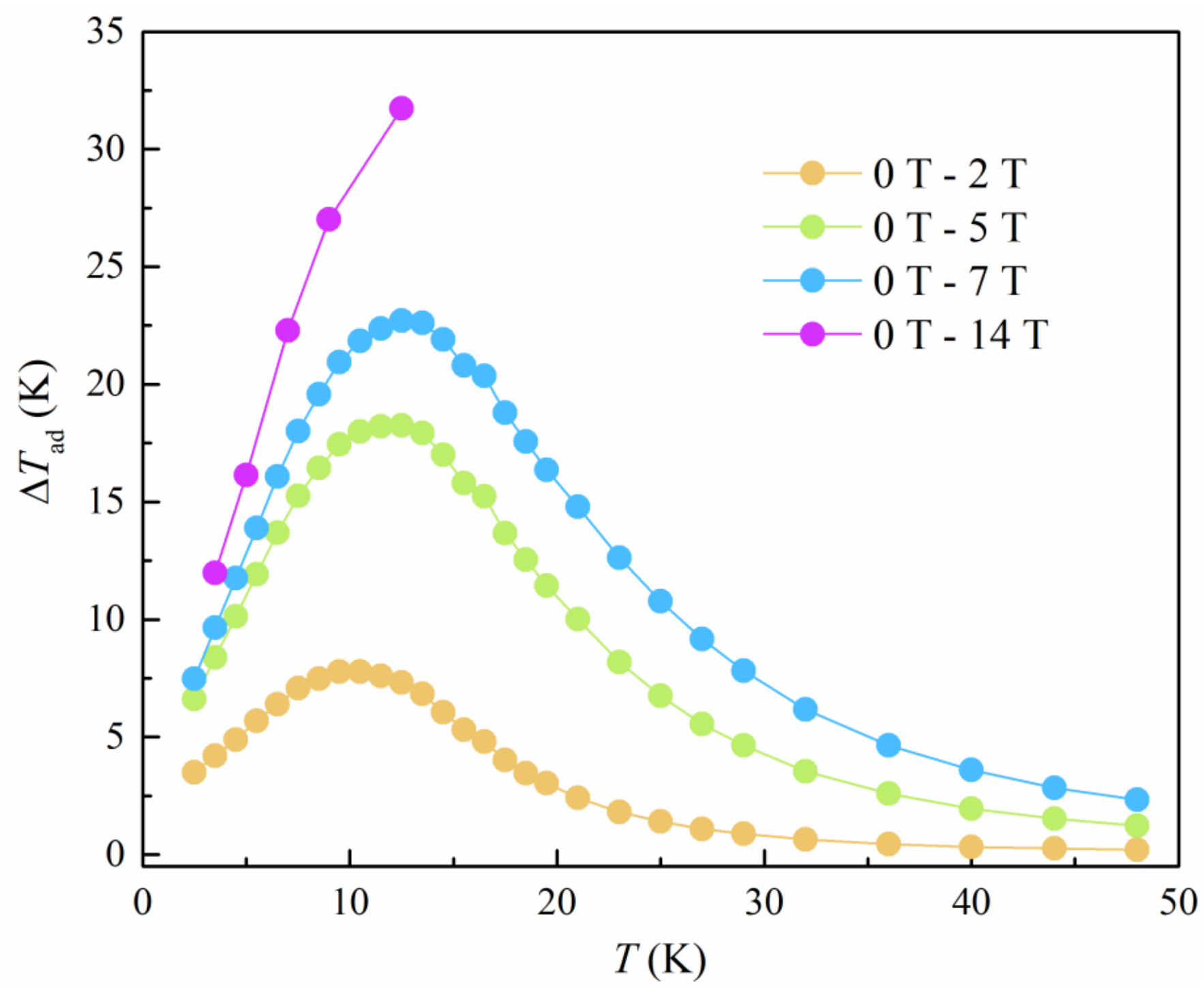

**Fig. 8.** Temperature dependence of adiabatic temperature change during the magnetic field change.

Having established the gravimetric and volumetric magnetic entropy changes as well as the calculated adiabatic temperature change of $Gd_2TeO_6$, it is useful to compare these parameters with those reported for other Gd based and related cryogenic magnetocaloric materials. A summary of the relevant magnetocaloric parameters is presented in **Table- I.**

The maximum gravimetric entropy change of 54.81 J kg$^{-1}$K$^{-1}$ obtained for $Gd_2TeO_6$ under a 0-7 T field change is comparable with the values reported for several recently investigated Gd based materials. For example, $Gd_3TeBO_9$, $Gd_3BWO_9$, and $Gd_2Si_2O_7$ exhibit reported $\Delta S_M$ values of 57.44, 54.80, and 58.81 J kg$^{-1}$ K$^{-1}$, respectively [44,45,49]. The value for $Gd_2TeO_6$ is also higher than those reported for $Gd_2CuTiO_6$, $Gd_2MgTiO_6$, $GdCrTiO_5$, Gd2WO6, and GGG, which exhibit values of approximately 51.4, 46.21, 36, 38.4, and 38.4 J kg$^{-1}$ K$^{-1}$, respectively [29,50–52,54].

The comparison with the commercially established $Gd_3Ga_5O_{12}$ (GGG) is particularly relevant. Under a 0-7 T field change, GGG exhibits a reported maximum $\Delta S_M$ of approximately 38.4 J kg$^{-1}$ K$^{-1}$, whereas $Gd_2TeO_6$ exhibits 54.81 J kg$^{-1}$ K$^{-1}$. Thus, the gravimetric entropy change of $Gd_2TeO_6$ is approximately 42.7 % higher than the reported value for GGG under the corresponding field change. The volumetric entropy change provides an additional basis for comparison. $Gd_2TeO_6$ exhibits a maximum of 400.5 mJ cm$^{-3}$ K$^{-1}$ at 7 T. This value is approximately 6.5 %, 47.8 %, and 5.9 % higher than the reported values for $GdPO_4$, $Gd_3Ga_5O_{12}$, and $Gd_2CuTiO_6$, respectively [47,50,52]. It is also comparable to the value reported for $Gd_3TeBO_9$ (408 mJ cm$^{-3}$ K$^{-1}$) [44]. On the other hand, higher volumetric entropy changes have been reported for $Gd_{9.33}[SiO_4]_6O_2$, and $Gd_5BSi_2O_{13}$, with reported values of approximately 430, and 461 mJ cm$^{-3}$ K$^{-1}$, respectively [12,46]. The low-field magnetocaloric response of $Gd_2TeO_6$ is also noteworthy. At 2 T, it exhibits $\Delta S_M$ = 21.40 J kg$^{-1}$ K$^{-1}$ and $\Delta S_V$ = 156.31 mJ cm$^{-3}$ K$^{-1}$, which are approximately 4.9 % and 7.8 % higher, respectively, than the corresponding values reported for GGG [53]. A particularly relevant recent comparison is provided by the geometrically frustrated $LiCaGd_5(BO_3)_6$ oxide, which exhibits a maximum of 40.8 J kg$^{-1}$ K$^{-1}$ under a low field change of 0-2 T, accompanied by a volumetric entropy change of approximately 118 mJ cm$^{-3}$ K$^{-1}$ [35]. Under the same 0-2 T field change, $Gd_2TeO_6$ therefore exhibits a volumetric entropy change approximately 32.5 % higher than $LiCaGd_5(BO_3)_6$. Moreover, the advantage becomes more pronounced at higher fields: at 7 T, the $\Delta S_V$ of $Gd_2TeO_6$ is approximately 126 % higher than that of $LiCaGd_5(BO_3)_6$. These comparisons demonstrate that $Gd_2TeO_6$ combines a high volumetric entropy density with a substantial low-field response, retaining a strong magnetocaloric performance even under relatively modest magnetic-field changes.

The calculated adiabatic temperature change can also be compared with reported values. It is worth noting that $\Delta T_{ad}$ can be determined either directly through adiabatic demagnetization refrigeration (ADR) measurements or indirectly from isentropic curves constructed from the magnetic entropy. In the present work, was estimated indirectly from the magnetization and heat-capacity measurements using Eq. (9). As summarized in **Table- I.**, reported values for $Gd_3BWO_9$, $Gd_{9.33}[SiO_4]_6O_2$, $GdFeO_3$, and $GdCrTiO_5$ are approximately 9.16 , 23.8, ≈ 21, and 15.5 K, respectively [12,45,54,55]. The calculated of 22.7 K at 7 T for $Gd_2TeO_6$ therefore lies within the range reported for several strongly magnetocaloric Gd-based compounds. It is lower than the reported value of 24.6 K for $GdPO_4$, while being comparable to the approximately 21 K reported for $Gd_2MoO_6$ under a similar field change [33,47]. Notably, a recent report on $Gd_2CuTiO_6$ gives a $\Delta T_{ad}$ of 22.6 K for a larger field change of 10 T, whereas $Gd_2TeO_6$ reaches a comparable value of 22.7 K for a 0-7 T field change [50]. This comparison highlights the substantial adiabatic temperature response of $Gd_2TeO_6$, particularly considering that it is calculated $\Delta T_{ad}$ is comparable to values reported for related Gd-based refrigerants despite being obtained under a smaller magnetic-field change. At 14 T, $Gd_2TeO_6$ exhibits a calculated $\Delta T_{ad}$ of approximately 31.7 K at 12.5 K.

Overall, the comparison demonstrates that $Gd_2TeO_6$ combines a large low-field magnetic entropy change, high gravimetric and volumetric entropy changes at 7 T, and a substantial calculated adiabatic temperature change. In particular, the simultaneous observation of 21.40 J kg$^{-1}$ K$^{-1}$ at 2 T, 54.81 J kg$^{-1}$ K$^{-1}$ and 400.5 mJ cm$^{-3}$ K$^{-1}$ at 7 T, together with a calculated $\Delta T_{ad}$ of 22.7 K at 7 T, establishes the strong cryogenic magnetocaloric response of $Gd_2TeO_6$. The near-theoretical entropy change of 63.51 J kg$^{-1}$ K$^{-1}$ at 14 T further indicates that a large fraction of the available $Gd^{3+}$ magnetic entropy can be accessed by magnetic-field-induced polarization.

**Table- I.** Comparison of the magnetocaloric properties of $Gd_2TeO_6$ with selected Gd-based and related materials. $^{\dagger}$represents $\Delta T_{ad}$ determined by direct ADR measurements,$^{*}$represents value calculated using the reported density, and NR represents not reported.

| **Material** | **Field change** | ***T* @ -$\Delta S_M^{max}$ (K)** | **-$\Delta S_M^{max}$ (J kg$^{-1}$ K$^{-1}$)** | **-$\Delta S_V^{max}$ (mJ cm$^{-3}$ K$^{-1}$)** | **$\Delta T_{ad}$ (K)** |
|---|---|---|---|---|---|
| **$Gd_2TeO_6$ (this work)** | **0-7 T** | **2.5 K** | **54.81** | **400.5** | **22.7** |
| $Gd_3TeBO_9$ [44] | 0-7 T | 2.6 | 57.44 | 408 | NR |
| $Gd_3BWO_9$ [45] | 0-7 T | 2 | 54.80 | NR | 9.16 |
| $Gd_{9.33}[SiO_4]_6O_2$ [12] | 0-7 T | 2 | 66.6 | 430 | 23.8 |
| $Gd_5BSi_2O_{13}$ [46] | 0-7 T | 3 | 67 | 461 | NR |
| $GdPO_4$ [47] | 0-7 T | 2.1 | 62 | 357.8 | 24.6 |
| $GdVO_4$ [48] | 0-7 T | ≈ 3 | 47.90 | NR | 22.39 Iso-entropic |
| $Gd_2Si_2O_7$ [49] | 0-7 T | 3 | 58.81 | NR | NR |
| $Gd_2CuTiO_6$ [50] | 0-7 T | ≈ 3 | 51.4 | 378.2 | 22.6 (0 T – 10 T) |
| $Gd_2MgTiO_6$ [51] | 0-7 T | 3.3 | 46.21 | NR | NR |
| $Gd_3Ga_5O_{12}$ (GGG) [52,53] | 0-7 T | ≈ 2.0 | 38.4 | 271 | 1.68 K for 4 T → 0 T from 2 K$^{\dagger}$ |
| $GdCrTiO_5$ [54] | 0-7 T | ≈ 2.0 | 36 | NR | 15.5 |
| $GdFeO_3$ [55] | 0-7 T | ≈ 3 | 44 | NR | ≈ 21 |
| $GdScO_3$ [56] | 0-7 T | ≈ 2.5 | 50.77 | NR | NR |
| $Gd_2WO_6$ [29] | 0-7 T | 4 | 34.8 | NR | NR |
| $Gd_2MoO_6$ [33] | 0-7 T | ≈ 3 K | 36.7 | NR | ≈ 21 |
| $Gd_2B_2MoO_9$ [16] | 0-5 T | ≈ 1 | 55 | 316.8$^{*}$ | 1.84 K for 9 T → 1.04 T from 2 K$^{\dagger}$ |
| $LiCaGd_5(BO_3)_6$ [35] | 0-7 T | ≈ 1.6 | ≈ 60 | ≈ 177 | NR |
| $K_3Gd_5(PO_4)_6$ [57] | 0-5 T | 2.5 | 43.7 | NR | 3.63 K for 4 T → 0 T from 4 K$^{\dagger}$ |
| $Eu_2B_2O_5$ [58] | 0-5 T | 2.7 | 46.9 | NR | 13.3 (0 – 5 T) |

## 4. Conclusion

In summary, we have systematically investigated the structural, magnetic, thermodynamic, and magnetocaloric properties of polycrystalline $Gd_2TeO_6$, complemented by first-principles calculations. Powder X-ray diffraction confirms the formation of single-phase orthorhombic $Gd_2TeO_6$ with space group $P2_12_12_1$, containing two crystallographically distinct seven-coordinated Gd sites. Magnetic measurements reveal weak antiferromagnetic interactions, as indicated by the negative Curie-Weiss temperature of -1.25 K, while no conventional long-range magnetic ordering is observed above 2 K. The magnetization reaches approximately 13.6 $\mu_B$ at 7 T at 2 K, close to the theoretical saturation moment, with negligible hysteresis, indicating a highly reversible field dependence.

$Gd_2TeO_6$ exhibits a substantial low-temperature magnetocaloric response, with $-\Delta S_M$ values of 21.40, 48.51 and 54.81 J kg$^{-1}$ K$^{-1}$ for field changes of 0-2, 0-5, and 0-7 T, respectively. The corresponding volumetric entropy change reaches 400.50 m J cm$^{-3}$ for 0-7 T. Extending the magnetization measurements to 14 T yields a maximum measured $-\Delta S_M$ of 63.51 J kg$^{-1}$ K$^{-1}$, approaching the theoretical magnetic entropy of 64.25 J kg$^{-1}$ K$^{-1}$ expected for two $Gd^{3+}$ ions per formula unit. The field-dependent heat-capacity measurements provide further evidence for field-induced redistribution of magnetic entropy, with agreement between the heat-capacity and magnetization derived entropy changes at 5 and 7 T. An indirect evaluation gives a maximum $\Delta T_{ad}$ of approximately 22.7 K for a 0-7 T field change.

The first-principles DFT calculations reveal closely competing antiferromagnetic configurations in $Gd_2TeO_6$. GGA favors an arrangement with alternating spin alignment within both crystallographically distinct Gd sublattices, whereas GGA+ $U$ stabilizes a configuration in which the moments within each Gd sublattice are parallel and the two groups are mutually antiparallel. The latter arrangement remains energetically favored upon variation of the k-point mesh and Hubbard $U$, indicating a relatively narrow magnetic-energy landscape within the investigated collinear configurations. Further measurements below 2 K and direct or quasi-adiabatic refrigeration experiments will be important for establishing the ultimate low-temperature magnetic and magnetocaloric performance. Overall, the combination of large magnetic entropy, weak magnetic interaction scale, negligible hysteresis, and substantial gravimetric and volumetric magnetocaloric responses makes $Gd_2TeO_6$ a promising candidate for further investigation as a cryogenic magnetic refrigerant.

**Acknowledgements:**

Abhijit Nayak thanks the NIT, Andhra Pradesh, for financial support and J K Murthy thanls DST-SERB sponsored Core Research Grant, India, for financial support under Grant No. CRG/20211005933. D.M. acknowledges the funding from the Technical Research Center, Department of Science and Technology, Government of India (Grant No. AI/1/62/IACS/2015) and the India-Russia Joint Research Call, Department of Science and Technology, Government of India (Grant No. DST/IC/RSF/2024/542).

# Supplementary Information (SI)

**Table -SI:** Atomic coordinates of Gd, Te, and O as obtained from the Rietveld refinement.

| | Atomic coordinates |
|---|---|
| Gd1 (x) | 0.0354(16) |
| y | 0.2199(7) |
| z | 0.1364(6) |
| Gd2 (x) | 0.019(2) |
| y | 0.1052(6) |
| z | 0.5403(6) |
| Te (x) | 0.5387(17) |
| y | 0.0051(7) |
| z | 0.3535(7) |
| O1 (x) | 0.311(10) |
| y | 0.370(9) |
| z | 0.263(6) |
| O2 (x) | 0.319(9) |
| y | 0.048(8) |
| z | 0.209(6) |
| O3(x) | 0.408(11) |
| y | 0.163(6) |
| z | 0.491(5) |
| O4(x) | 0.286(9) |
| y | 0.355(8) |
| z | 0.671(7) |
| O5 (x) | 0.243(13) |
| y | 0.119(7) |
| z | 0.929(5) |
| O6 (x) | 0.252(12) |
| y | 0.414(9) |
| z | 0.006(6) |

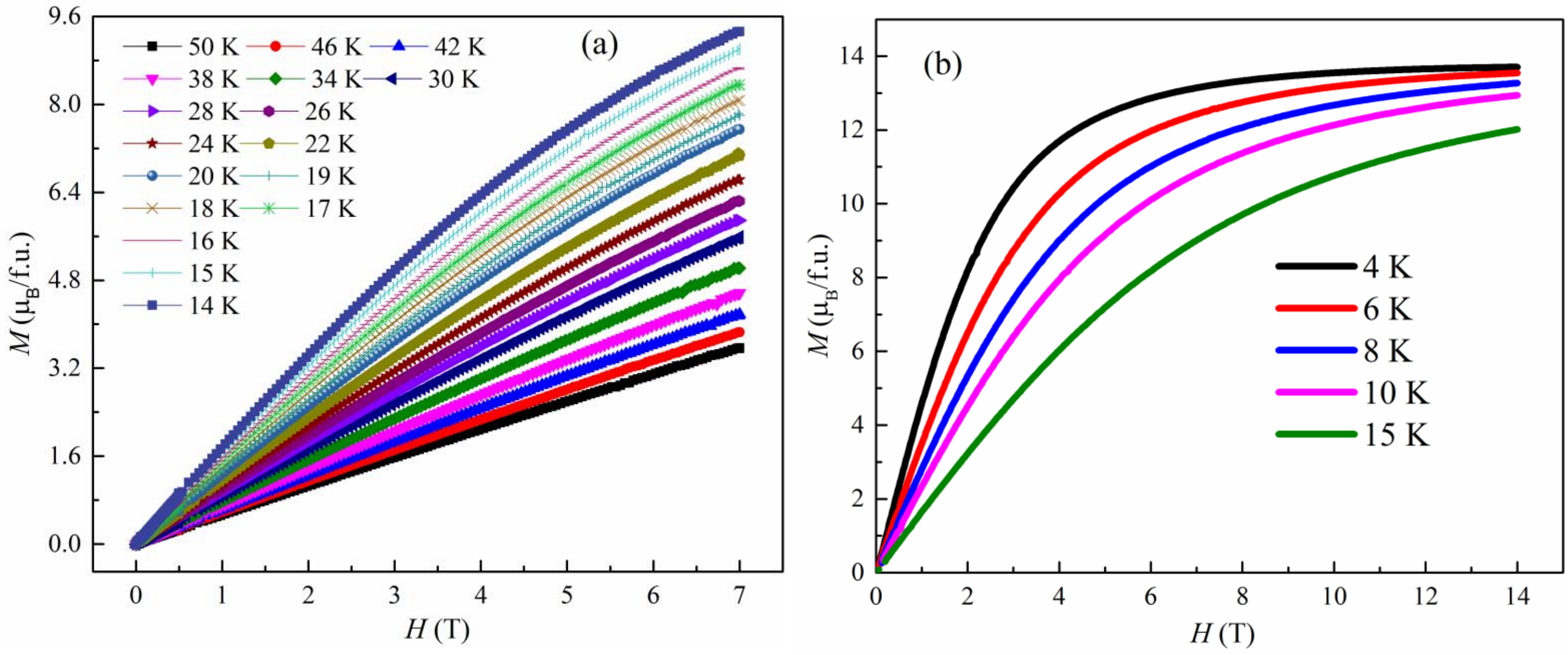

**Fig. S1.** Field dependence of magnetization of $Gd_2TeO_6$ at different temperatures, **(a)** with the field range of 0 T- 7 T, and **(b)** 0 T - 14 T.

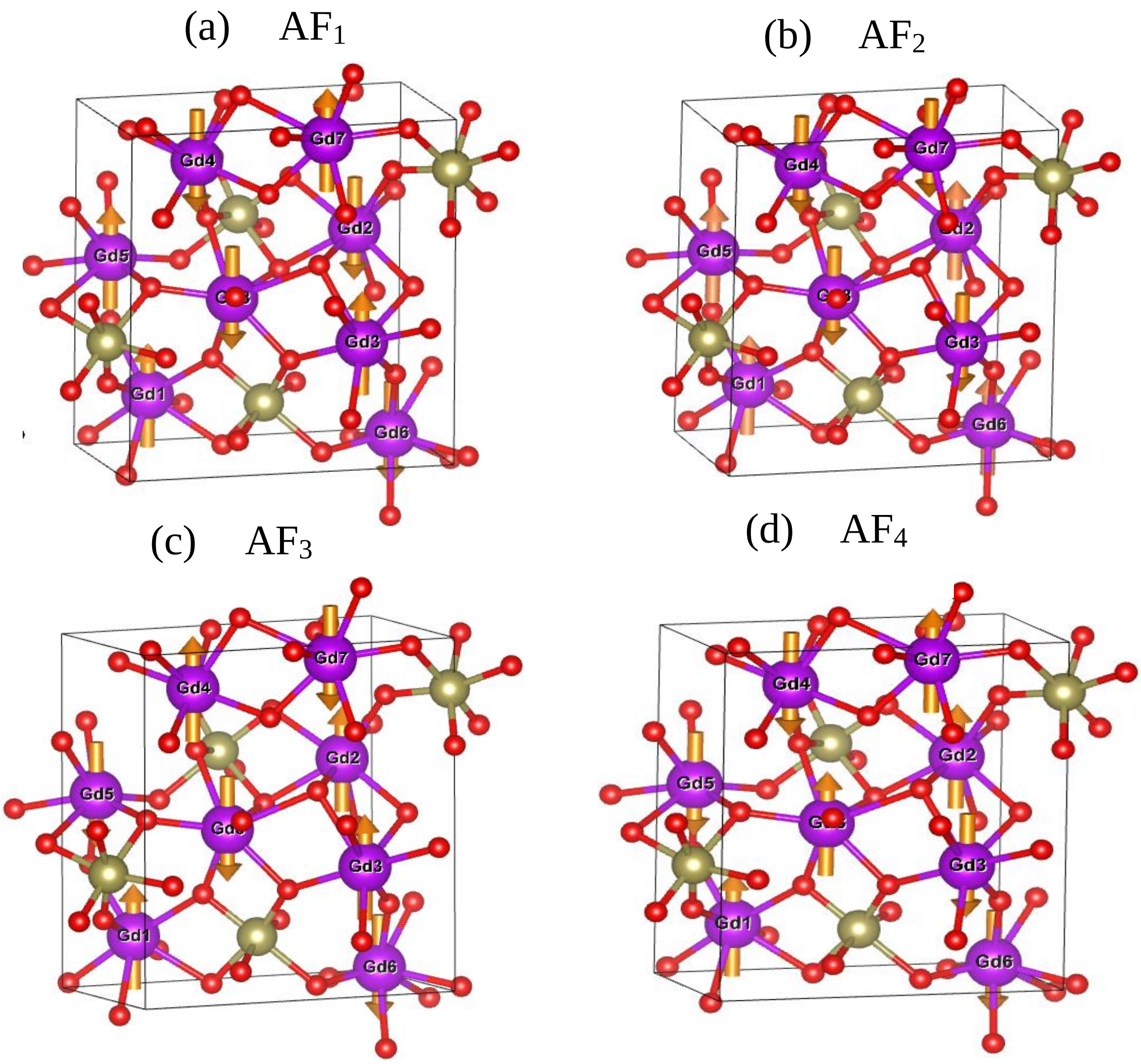

**Fig. S2.** Schematic representations of the spin configurations of the eight Gd atoms considered in the DFT calculations: **(a)** $AF_1$, **(b)** $AF_2$, **(c)** $AF_3$, and **(d)** $AF_4$.

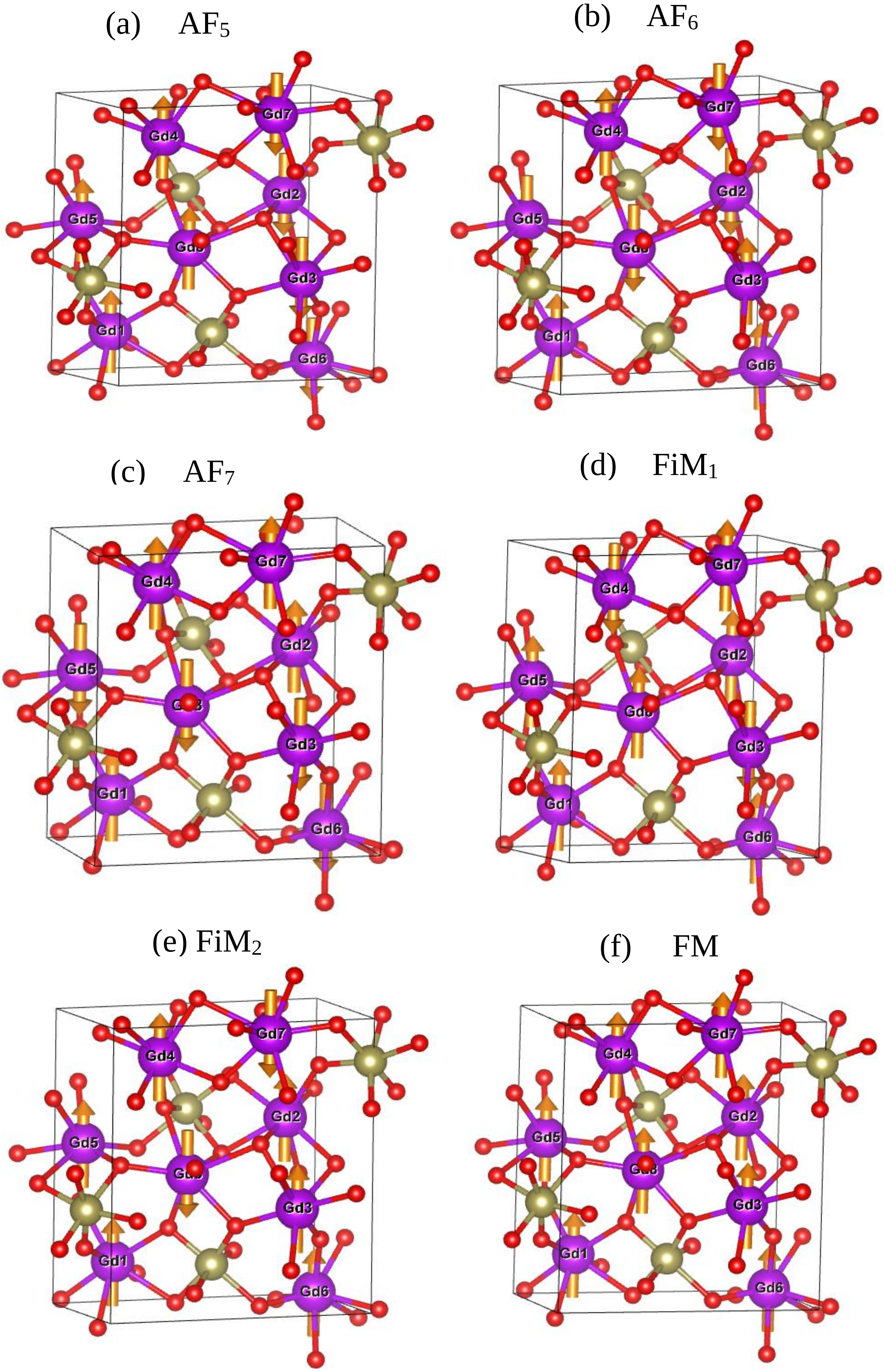

**Fig. S3.** Schematic representations of the spin configurations of the eight Gd atoms considered in the DFT calculations: **(a)** $AF_5$, **(b)** $AF_6$, **(c)** $AF_7$, **(d)** $FiM_1$, **(e)** $FiM_2$, and **(f)** FM.

**Table- SII:** Total and relative energy a of different spin structures obtained with the GGA approach with energy cutoff of $10^{-6}$ eV and KPOINT of 8*6*5

| Spin Configuration | Total Energy (eV) | Energy difference (ΔE) (eV) |
|---|---|---|
| $AF_1$ | -334.72354001 | -0.00022140 |
| $AF_2$ | -334.72120600 | -0.00255541 |
| $AF_3$ | -334.72318816 | -0.00057325 |
| $AF_4$ | -334.72376141 | 0 |
| $AF_5$ | -334.72108366 | -0.00267775 |
| $AF_6$ | -334.72174792 | -0.00201349 |
| $AF_7$ | -334.72172756 | -0.00203385 |
| $FiM_1$ | -334.72072290 | -0.00303851 |
| $FiM_2$ | -333.69511910 | -0.02864231 |
| FM | -333.98491836 | -0.73884305 |

**Table- SIII:** Total and relative energy a of $AF_1$, $AF_3$, and $AF_4$ spin structures within the GGA approach with the energy cutoff of $10^{-8}$ eV.

| | Total Energy (eV) | Energy difference (eV) |
|---|---|---|
| $AF_1$ | -334.72354024 | -0.00022137 |
| $AF_3$ | -334.72318835 | -0.00057326 |
| $AF_4$ | -334.72376161 | 0 |

**Table- SIV:** Total and relative energies of different spin structures obtained with GGA+$U$ formalism with $U$ = 6 eV, with the energy cutoff of $10^{-8}$ eV and k-mesh of 6*4*3

| Spin Configuration | Total Energy (eV) | Energy difference (ΔE) (eV) |
|---|---|---|
| $AF_1$ | -333.69806346 | 0.00056038 |
| $AF_2$ | -333.69543389 | 0.00318995 |
| $AF_3$ | -333.69862384 | 0 |
| $AF_4$ | -333.69733928 | 0.00128456 |
| $AF_5$ | -333.69389346 | 0.00473038 |
| $AF_6$ | -333.69611020 | 0.00251364 |
| $AF_7$ | -333.69606245 | 0.00256139 |
| $FiM_1$ | -333.69463007 | 0.00399377 |
| $FiM_2$ | -333.69511816 | 0.00350568 |
| FM | -333.68812015 | 0.01050369 |

**Table- SV:** Total and relative energy ($E$AF$_3$ - $E$AF1) a of $AF_1$ and $AF_3$ spin structures obtained with the k-mesh of 8*6*5, energy cutoff of $10^{-8}$ eV, with GGA+$U$ formalism with $U$ = 5, 6, 7, and 8 eV, respectively.

| Spin Configuration | $U$ = 5 eV | $U$ = 6 eV | $U$ = 7 eV | $U$ = 8 eV |
|---|---|---|---|---|
| $AF_1$ | -333.59652651 | -333.69805225 | -333.91721580 | -334.26781164 |
| $AF_3$ | -333.59709215 | -333.69865831 | -333.91782831 | -334.26839402 |
| (Δ$E$) eV | -0.00056564 | -0.00060606 | -0.00061251 | -0.00058238 |